\documentclass{llncs}

\pdfoutput=1

\usepackage[greek,english]{babel} 
\usepackage{makeidx} 
\usepackage{ulem}
\usepackage{longtable}
\usepackage{amssymb}
\usepackage{multirow}
\usepackage{enumitem}

\usepackage{lscape}
\usepackage{subfigure}
\usepackage{caption}
\usepackage{graphicx}
\usepackage{float}

\usepackage{textcomp}

\newcommand{\sppbench}{\(\mathrm{SP^{2}Bench}\)}

\begin{document}

\title{Geographica: A Benchmark \\ for Geospatial RDF Stores
\thanks{This work was supported in part by the European Commission project
TELEIOS (257662)}
}

\author{George Garbis \and Kostis Kyzirakos \and Manolis Koubarakis} 

\institute{National and Kapodistrian University of Athens, Greece\\
\{ggarbis,kk,koubarak\}@di.uoa.gr}

\maketitle         

\begin{abstract}
Geospatial extensions of SPARQL like GeoSPARQL and \linebreak stSPARQL have
recently been defined and corresponding geospatial RDF stores have been implemented. However,
there is no widely used benchmark for evaluating geospatial RDF stores which
takes into account recent advances to the state of the art in this area. In this
paper, we develop a benchmark, called Geographica, which uses both real-world
and synthetic data to test the offered functionality and the performance of some
prominent geospatial RDF stores.
\end{abstract}

\section{Introduction}
\label{sec:introduction}
The Web of data has recently started being populated with geospatial data and
geospatial extensions of SPARQL, like GeoSPARQL and stSPARQL, have been defined.
GeoSPARQL \cite{ogc-geosparql} is a recently proposed OGC standard for a
SPARQL-based query language for geospatial data expressed in RDF.
GeoSPARQL defines a vocabulary (classes, properties, and extension functions)
that can be used in RDF graphs and SPARQL queries to represent and query
geographic features with vector geometries.
stSPARQL \cite{iswc2012-strabon,eswc-temporal} is an extension of SPARQL 1.1
developed by our group for representing and querying geospatial data that change
over time.
Similarly to GeoSPARQL, the geospatial part of stSPARQL defines datatypes that can be used for
representing in RDF the serializations of vector geometries encoded according
to the widely adopted OGC standards Well Known Text (WKT) and Geography Markup
Language (GML). stSPARQL and GeoSPARQL define extension functions from the
OGC standard ``OpenGIS Simple Feature Access'' (OGC-SFA) that can be
used by the users for manipulating vector geometries.

In parallel with the appearance of GeoSPARQL and stSPARQL, researchers have
implemented geospatial RDF stores that support these SPARQL extensions (our own
system Strabon\footnote{\url{http://strabon.di.uoa.gr/}},
Parliament\footnote{\url{http://parliament.semwebcentral.org/}} and
uSeekM\footnote{\url{http://dev.opensahara.com/projects/useekm/}}). Typically,
this has been done by extending existing RDF stores that had no geospatial
functionalities (e.g., Sesame) and by relying in state of the art
spatially-enabled RDBMS (e.g., PostGIS) for the storage and querying of
geometries. One reason that this approach has been successful is that the
relational realization of the OGC-SFA standard has been widely adopted by many
RDBMS  for storing and manipulating vector geometries. The state of the art in
this area is summarized in the survey paper \cite{rr-survey}.

The above advances to the state of the art in query languages and implemented
systems has not so far been matched with much work on the evaluation and
benchmarking of implemented geospatial RDF stores. Although there are
various benchmarks for spatially-enabled
RDBMS~\cite{sequoia,Patel97buildinga,alacarte,vespa,jackpine,dynamark}, there
is only one paper in the literature that proposes a benchmark for geospatial
data expressed in RDF \cite{kolas}.
However, since this work has preceded the proposal of GeoSPARQL and stSPARQL,
it does not cover much of the features available in these languages. For
example, only point and rectangle geometries are used in the data and only two
topological functions and two non-topological functions are considered, while
metric spatial functions and spatial aggregates are not discussed.
Similarly, only the geospatial RDF store SPAUK, which is a precursor to
Parliament, has been evaluated using the benchmark. Finally, \cite{kolas} uses
a synthetic workload only and does not consider any linked geospatial datasets
such as the ones that are available in the LOD cloud today.

In this paper we go significantly beyond \cite{kolas} and develop a benchmark,
that can be used for the evaluation of the new generation of RDF stores
supporting the query languages GeoSPARQL and stSPARQL. Our benchmark,
nick-named Geographica,
\footnote{Geographica
(Greek: \greektext Gewgrafik'a\latintext)
is a 17-volume encyclopedia of geographical
knowledge written by the greek geographer, philosopher and historian Strabon 
(Greek: \greektext Str'abwn\latintext)
in 7 BC. (\url{http://en.wikipedia.org/wiki/Geographica})} 
is composed by two workloads with their associated datasets and queries: a
\textit{real-world} workload based on publicly available linked data sets and a
\textit{synthetic} workload. The real-world workload uses publicly available
linked geospatial data, covering a wide range of geometry types (e.g., points, lines,
polygons). To define this workload, we follow the approach of the benchmark
Jackpine \cite{jackpine} and we define a micro benchmark and a macro benchmark.
The micro benchmark tests primitive spatial functions. We check the spatial
component of a system with queries that use non-topological functions, spatial
selections, spatial joins and spatial aggregate functions. In the macro
benchmark we test the performance of the selected RDF stores in typical
application scenarios like reverse geocoding, map search and browsing, and a
real-world use case from the Earth Observation domain.
In the second workload of Geographica we use a generator that produces
synthetic datasets of various sizes and generates queries of varying thematic
and spatial selectivity. In this way, we can perform the evaluation of
geospatial RDF stores in a controlled environment. In this part we follow the
rationale of  earlier papers \cite{vespa,iswc2012-strabon,brodtRDF3Xgermanoi}.
For reasons of reproducibility, both workloads are publicly
available\footnote{\url{http://geographica.di.uoa.gr/}}.

We chose to test the systems Strabon, Parliament and uSeekM. To the best
of our knowledge, these systems are the only ones that currently provide
support for a rich subset of GeoSPARQL and stSPARQL. Other RDF stores like
OpenLink Virtuoso, OWLIM and AllegroGraph, allow only the representation of
point geometries and provide support for a few geospatial
functions~\cite{rr-survey}. The limited functionality provided by these systems
did not allow us to include them in our experiments. However, these systems can
be evaluated in the future if one wishes to test how well they perform for the
limited functionalities that they offer.

The rest of the paper is organized as follows. Section~\ref{sec:relatedwork}
presents previous related work. The benchmark and its results are described in
Sections~\ref{sec:benchmark} and \ref{sec:benchmarkresults}, respectively and 
general conclusions and future work  are discussed in
Section~\ref{sec:conclusions}.

\section{Related Work}
\label{sec:relatedwork}
In this section we discuss the most important benchmarks that are relevant
to Geographica. We first present well-known benchmarks for SPARQL query 
processing, then benchmarks from the area of spatial relational databases and, finally,
the only available benchmark for querying linked geospatial data.
 
\paragraph{Benchmarks for SPARQL query processing.} Four well-known benchmarks for SPARQL querying are the 
Lehigh University Benchmark (LUBM)~\cite{lubm}, the Berlin SPARQL Benchmark 
(BSBM)~\cite{bsbm}, the \sppbench{} SPARQL Performance Benchmark~\cite{sppbench} 
and the DBPedia SPARQL Benchmark (DBPSB)~\cite{dbpsb}. LUBM, BSBM and \sppbench{}
create a 
synthetic dataset based on a use case scenario and define some queries covering
a spectrum of SPARQL characteristics. 
For example, the synthetic dataset of \sppbench{}
resembles the original publications dataset of DBLP while the dataset of LUBM describes the university
domain. 
The creators of DBPSB take a different approach. They propose a benchmark creation methodology based on real-world data
and query logs. 
The proposed methodology is used in~\cite{dbpsb} to create a benchmark based 
on DBPedia data and query-logs.

A recent activity in the area of benchmarking of RDF databases is the European 
project LDBC\footnote{\url{http://www.ldbc.eu/}}
which brings together researchers form databases and the Semantic Web, as well
as RDF and graph database technology 
vendors to develop benchmarks for RDF and graph databases.

\paragraph{Benchmarks for spatial relational databases.}
One of the first benchmarks for spatial relational databases has been the SEQUOIA
benchmark~\cite{sequoia}
which focuses on Earth Science use cases. In order for its results to be 
representative of Earth Sciences use cases, SEQUOIA uses real-world data
(satellite raster data, point locations of geographic features, land use/land cover polygons and data about
drainage networks covering the area of USA) and real-world queries. Its 
queries cover tasks like data loading, raster data management, filtering based 
on spatial and non-spatial criteria, spatial joins, and path computations over graphs.  
The SEQUOIA benchmark has been extended in~\cite{Patel97buildinga} to evaluate 
the geospatial DBMS Paradise. 
Two other well known benchmarks for spatial relational databases which use 
synthetic vector data are \'{A} La Carte~\cite{alacarte} and VESPA\cite{vespa}.
\'{A} La Carte uses a  dataset consisting only of rectangles
which are generated according to various statistical distributions 
and it has been used to compare the performance of different spatial join techniques.
VESPA~\cite{vespa} creates a more complex dataset with more geometry types
(polygons, lines and points) and it has been used to compare PostgreSQL with 
Rock \& Roll deductive object oriented database.
More recently, \cite{jackpine} has defined a more generic 
benchmark for spatial relational databases, called Jackpine. It 
includes two kinds of benchmarking, micro and macro. Micro benchmarking tests 
topological predicates and spatial analysis functions in 
isolation. 
Macro benchmarking defines six typical spatial data applications scenarios and
tests a number of queries based on them. 

\paragraph{Benchmarks for geospatial RDF stores.}
The only published benchmark for querying geospatial data encoded in RDF has 
been proposed by Kolas~\cite{kolas}. He extends LUBM to include spatial 
entities and to test the functionality of spatially  enabled RDF stores. 
LUBM queries are extended to cover four primary types of spatial queries, 
namely spatial location queries, spatial range queries, spatial join queries, 
nearest neighbor queries. Range queries aim to test cases of various selectivity, 
while spatial joins aims to test whether the query planner selects a good plan
by taking into account the selectivity of the spatial and ontological part 
of each query.

\section{The Benchmark Geographica}
\label{sec:benchmark}

	In this section we present our benchmark in detail. Section
\ref{sec:realworkload} presents its first part (the real-world workload) while
Section \ref{sec:syntheticworkload} presents the second part (the synthetic
workload).

	\subsection{Real-World Workload}
	\label{sec:realworkload}
		This workload aims at evaluating the efficiency of basic spatial 
		functions that a geospatial RDF store should offer. In addition, 
		this workload includes three typical application scenarios.
		\subsubsection{Datasets.}
		\label{sec:dataModel}
		\begin{table}[t]
\begin{center}
\begin{tabular}{|c|c|c|c|c|c|c|}
\hline
\textbf{Datasets} & \textbf{Size} & \textbf{Triples} & \textbf{\# of Points} & \textbf{\# of Lines} & \textbf{\# of Polygons} \\
\hline
 GAG & 33MB & 4K  & - & - & 325 \\ 
\hline
 CLC & 401MB & 630K  & - & - & 45K \\ 
\hline
 LGD (only ways) & 29MB & 150K & - & 12K & - \\ 
\hline
 GeoNames & 45MB & 400K  & 22K & - & - \\
\hline
 DBPedia & 89MB & 430K  & 8K & - & - \\
\hline 
 Hotspots & 90MB & 450K & - & - & 37K \\ 
\hline
\end{tabular}
\end{center}
\caption{Dataset characteristics}
\label{table:datasets}
\end{table}
In this section we describe the datasets that we use for the real-world workload.
We have datasets that play an important role in the 
Linked Open Data Cloud, such as the part of DBPedia and GeoNames referring to 
Greece, despite the fact that their spatial
information is limited to points. In addition we have part of the 
LinkedGeoData\footnote{\url{http://linkedgeodata.org/}} (LGD) 
dataset which 
has richer geospatial information from OpenStreetMap\footnote{\url{http://www.openstreetmap.org/}} 
about the road network and rivers of Greece. We also chose to use the Greek Administrative 
Geography\footnote{\url{http://www.linkedopendata.gr/dataset/greek-administrative-geography/}} (GAG) 
and the CORINE Land Use/Land Cover\footnote{\url{http://www.linkedopendata.gr/dataset/corine-land-cover-of-greece}} (CLC) 
dataset for Greece which have 
complex polygons. 
The CLC dataset is made available by the European Environmental Agency for the 
whole Europe and contains data regarding the land cover of European countries.
Both of these datasets with information about Greece have been published 
as linked data by us
in the context of the European project TELEIOS\footnote{\url{http://www.earthobservatory.eu/}}.
Finally, we include a dataset containing polygons that represent wild fire hotspots. 
This dataset has been produced by the National Observatory of Athens (NOA) in 
the context of project TELEIOS by processing
appropriate satellite images as described in~\cite{edbt2012}. 
Each dataset is loaded in a separate named graph so that each query access only the part
of the dataset that is needed.
Some important characteristics of the datasets used can be found in Table~\ref{table:datasets}.

		\subsubsection{Micro Benchmark.}
		\label{sec:micro-def}
		The micro benchmark aims at testing the efficiency of primitive spatial functions 
in state of the art geospatial RDF stores. Thus, we use simple SPARQL queries which
consist of one or two triple patterns and a spatial function. 
We start by checking simple spatial selections.
Next, we test more complex operations such as spatial joins.
We test spatial joins using the topological relations defined by stSPARQL~\cite{iswc2012-strabon} and the 
Geometry Topology component of GeoSPARQL. 
 
Apart from topological relations, we test non-topological functions as defined by stSPARQL
and the Geometry extension of GeoSPARQL as well. 
These functions (e.g., \texttt{strdf:area}, \texttt{geof:buffer}) calculate scalar values
 or construct a new geometry object.
The aggregate functions \texttt{strdf:extent}, and \texttt{strdf:union} of stSPARQL
are also tested by this benchmark. GeoSPARQL
does not define aggregate functions. 
We include aggregate functions in Geographica since present they are in all geospatial
RDBMS, and we found them very useful in EO applications in the context of the project TELEIOS.
A short description of queries used in the micro 
benchmark can be found in Table~\ref{table:microbenchmar1}.

		\begin{table}[t]
\begin{center}
\scriptsize
\begin{tabular}{|p{.08\textwidth}|p{.15\textwidth}|p{.77\textwidth}|}
\hline
\textbf{Query} & \textbf{Operation} & \textbf{Description} \\
\multicolumn{3}{|l|}{\textbf{Non-topological construct functions}} \\
\hline
Q1 &  Boundary & Construct the boundary of all polygons of CLC\\
\hline
Q2 & Envelope & Construct the envelope of all polygons of CLC\\
\hline
Q3 & Convex Hull & Construct the convex hull of all polygons of CLC\\
\hline
Q4 & Buffer & Construct the buffer of all points of GeoNames\\
\hline
Q5 & Buffer & Construct the buffer of all lines of LGD\\
\hline
Q6 & Area & Compute the area of all polygons of CLC\\
\hline
\multicolumn{3}{|l|}{\textbf{Spatial selections}} \\
\hline
Q7 & Equals & Find all lines of LGD that are spatially equal with a given line\\
\hline
Q8 & Equals & Find all polygons of GAG that are spatially equal a given  polygon\\
\hline
Q9 & Intersects & Find all lines of LGD that spatially intersect with a given polygon\\
\hline
Q10 & Intersects & Find all polygons of GAG that spatially intersect with a given line\\
\hline
Q11 & Overlaps & Find all polygons of GAG that spatially overlap with a given polygon\\
\hline
Q12 & Crosses & Find all lines of LGD that spatially cross a given line\\
\hline
Q13 & Within polygon & Find all points of GeoNames that are contained in a given polygon of GAG\\
\hline
Q14 & Within buffer of a point & Find all points of GeoNames that are contained in the buffer of a given point\\
\hline
Q15 & Near a point & Find all points of GeoNames that are within specific distance from a given point \\
\hline
Q16 & Disjoint & Find all points of GeoNames that are spatially disjoint of a given polygon\\
\hline
Q17 & Disjoint & Find all lines of LGD that are spatially disjoint of a given polygon\\
\hline
\multicolumn{3}{|l|}{\textbf{Spatial joins}} \\
\hline
Q18 & Equals & Find all points of GeoNames that are spatially equal with a point of DBPedia\\
\hline
Q19 & Intersects & Find all points of GeoNames that spatially intersect a line of LGD\\
\hline
Q20 & Intersects & Find all points of GeoNames that spatially intersect a polygon of GAG\\
\hline
Q21 & Intersects & Find all lines of LGD that spatially intersect a polygon of GAG\\
\hline
Q22 & Within & Find all points of GeoNames that are within a polygon of GAG\\
\hline
Q23 & Within & Find all lines of LGD that are within a polygon of GAG\\
\hline
Q24 & Within & Find all polygons of CLC that are within a polygon of GAG\\
\hline
Q25 & Crosses & Find all lines of LGD that spatially cross a polygon of GAG\\
\hline
Q26 & Touches & Find all polygons of GAG that spatially touch other polygons of GAG\\
\hline
Q27 & Overlaps & Find all polygons of CLC that spatially overlap polygons of GAG\\
\hline
\multicolumn{3}{|l|}{\textbf{Aggregate functions}} \\
\hline
Q28 & Extension & Construct the extension of all polygons of GAG \\
\hline
Q29 & Union & Construct the union of all polygons of GAG\\
\hline
\end{tabular}
\end{center}
\caption{Queries of the micro benchmark}
\label{table:microbenchmar1}
\end{table}

\begin{table}[htbp]
\begin{center}
\scriptsize
\begin{tabular}{|l|p{.90\textwidth}|}
\hline
\textbf{Query} & \textbf{Description} \\
\hline
\multicolumn{2}{|l|}{\textbf{Reverse Geocoding}} \\
\hline
RG1 & Find the city which is closest to a given point \\
\hline
RG2 & Find the street which is closest to a given point \\
\hline
\multicolumn{2}{|l|}{\textbf{Map Search and Browsing}} \\
\hline
MSB1 & Find POI satisfying some thematic criteria \\
\hline
MSB2 & Retrieve roads around a POI \\
\hline
MSB3 & Retrieve buildings around a POI \\
\hline\multicolumn{2}{|l|}{\textbf{Rapid Mapping for Fire Monitoring}} \\
\hline
RM1 & Find the land cover area of a bounding box  \\
\hline
RM2 & Find all primary roads contained in a bounding box \\
\hline
RM3 & Find all capitals of prefectures in a bounding box \\
\hline
RM4 & Find all municipality boundaries in bounding box \\
\hline
RM5 & Find all coniferous forests which are on fire \\
\hline
RM6 & Find road segments which may be damaged by fire \\
\hline
\end{tabular}
\end{center}
\caption{Queries of the macro benchmark}
\label{table:macrobenchmark}
\end{table}

		\subsubsection{Macro Benchmark.}
		\label{sec:macro-def}
		In the macro benchmark we aim to test the performance of the selected RDF 
stores in the following typical application scenarios: reverse geocoding, map
search and browsing, and two scenarios from the Earth Observation domain.
 
\paragraph{Reverse Geocoding.}
Reverse geocoding is the process of attributing a readable address or place name
to a given point. Thus, in this scenario, we pose SPARQL queries which sort
retrieved objects by their distance to the given point and select the first one.

\paragraph{Map Search and Browsing.}
This scenario demonstrates the queries that are typically used in Web-based mapping 
applications. A user first searches for points of interest based on thematic
criteria. Then, he/she selects a specific point and information about the area 
around it is retrieved (e.g., POI and roads).
 
\paragraph{Rapid Mapping for Wild Fire Monitoring.}
In this scenario we test queries which retrieve map layers for creating a 
map that can be used by decision makers tasked with the monitoring
of wild fires. This application has been studied in detail in project
TELEIOS\cite{edbt2012} and the scenario covers its core querying needs.
First, spatial 
selections are used to retrieve basic information of interest
(e.g., roads, administrative areas etc.). Second more complex information can 
be derived using spatial joins and non-topological functions. For example, a 
user may be interested in the segment of roads that may be damaged by fire. We
point out that this scenario is representative of many rapid mapping
tasks encountered in Earth Observation applications.
The queries of the macro benchmark can be found in Table~\ref{table:macrobenchmark}.

	\subsection{Synthetic Workload}
	  \label{sec:syntheticworkload}
		The synthetic workload of Geographica relies on a generator that produces synthetic
		datasets of various sizes and instantiates query templates that can produce
		queries with varying thematic and spatial selectivity. In this way, we can perform the
		evaluation of geospatial RDF stores in a controlled environment in
		order to monitor their performance with great precision.		
		\subsubsection{Datasets.}
		\label{sec:syntheticDataset}
		\begin{figure}[t]
 \begin{center}
  \subfigure[][\shortstack{\\ Ontology for Points of Interest}]
	  {\includegraphics[scale=0.35]{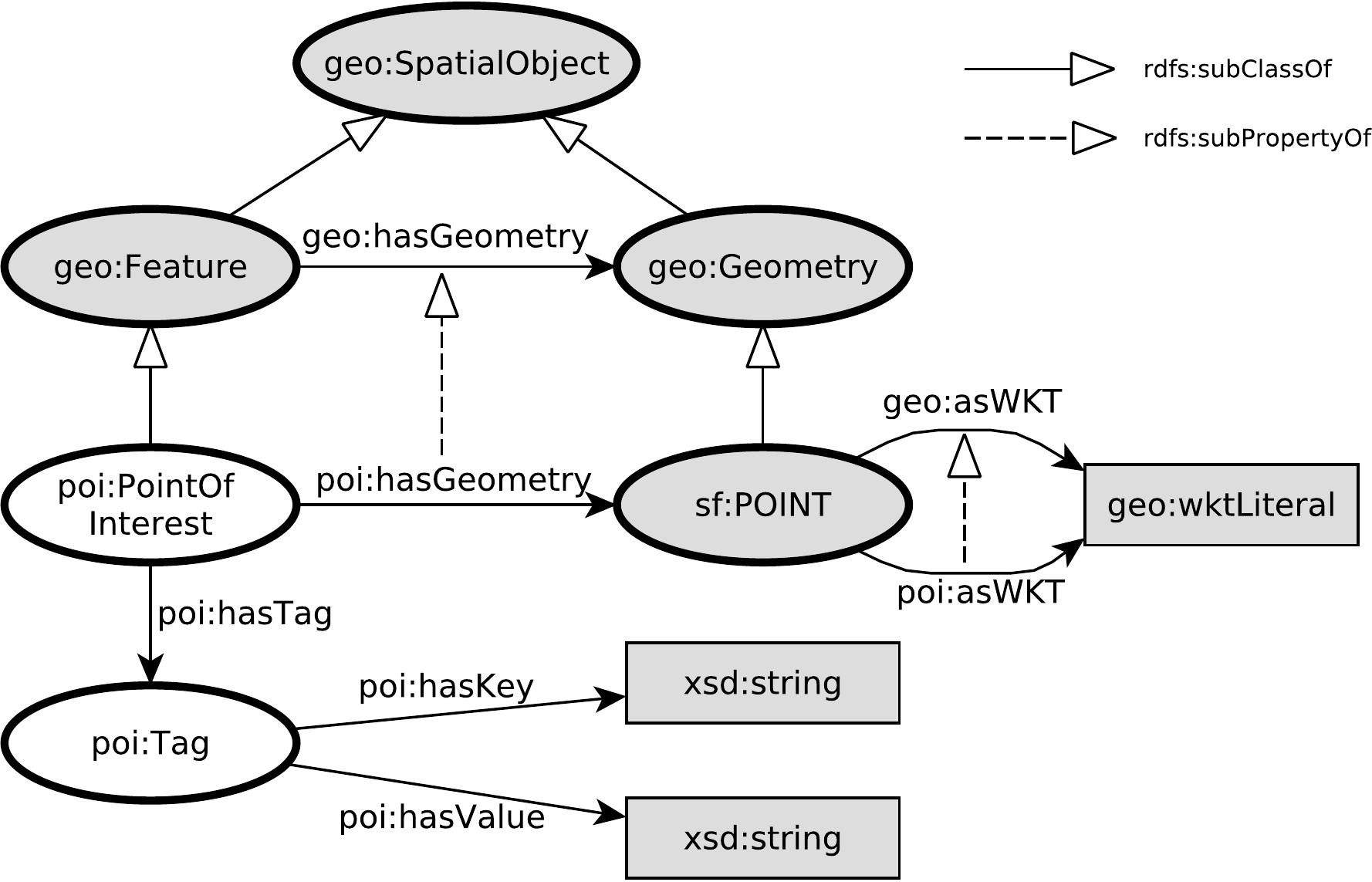}
	  \label{fig:sytheticDatasetOntology}}
	  \subfigure[][Visualization of the geometric part of the synthetic dataset]
	  {\includegraphics[scale=0.18]{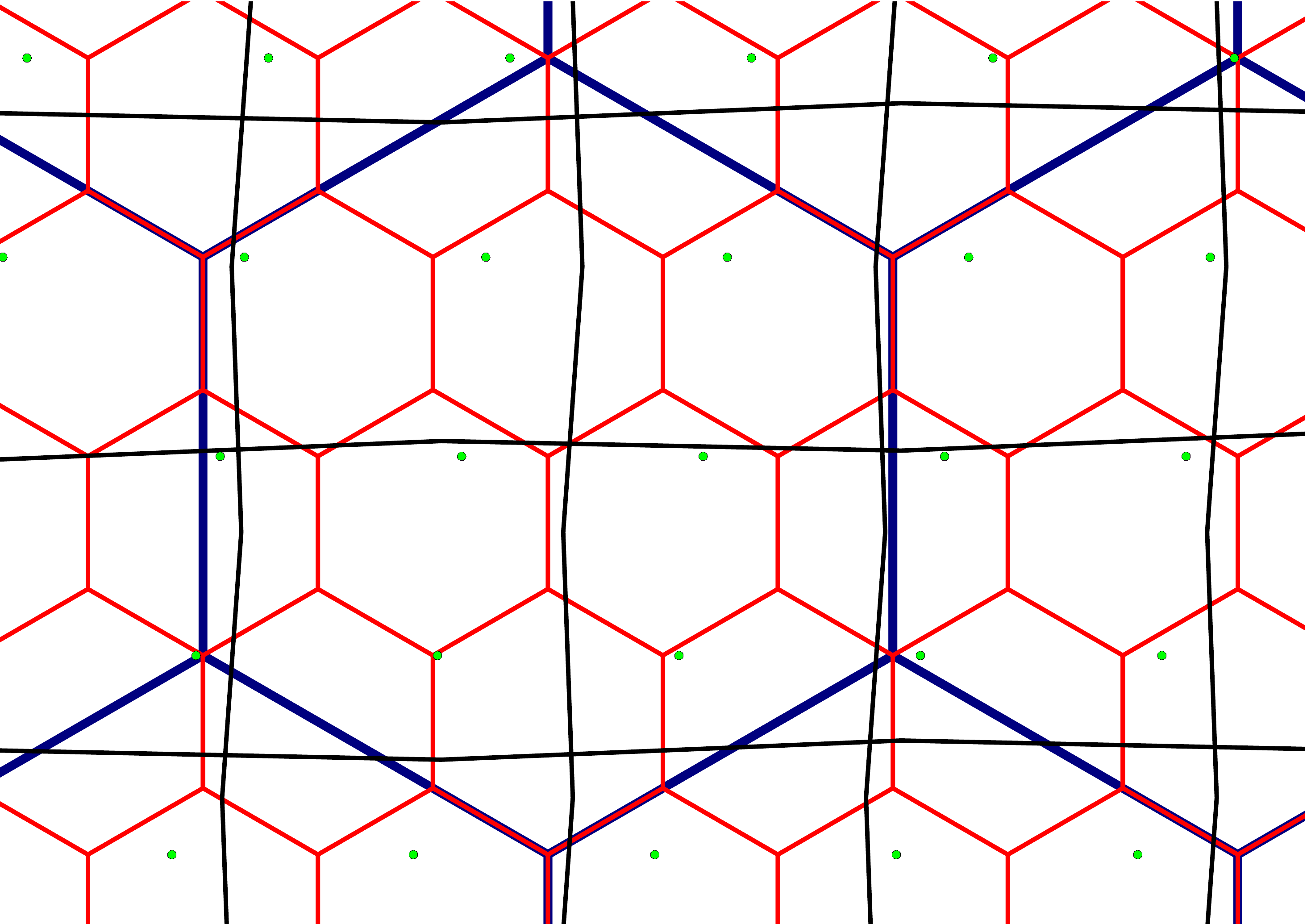}
	  \label{fig:sytheticDatasetVisualization}}
	\caption{Synthetic dataset}
	\label{fig:sytheticDataset}
\end{center}
\end{figure}

The workload generator produces synthetic datasets
of arbitrary size that resemble features on a map. As in VESPA~\cite{vespa},
the produced datasets model the following geographic features: states in a country,
land ownership, roads and points of interest.
For each dataset, we
developed a minimal
ontology\footnote{\url{http://geographica.di.uoa.gr/generator/\{ontology,
landOwnership, state, road, pointOfInterest\}}} that follows a general version
of the schema of OpenStreetMap and uses GeoSPARQL ontologies and vocabularies. 
In Figure~\ref{fig:sytheticDatasetOntology} we present the developed
ontology for representing points of interest only. 
As in \cite{brodtRDF3Xgermanoi,iswc2012-strabon}, every feature
(i.e., point of interest) is assigned a number of thematic tags each of which consists of
a key-value pair of strings. Each feature is tagged with key $1$, every
other feature with key $2$, every fourth feature with key $4$,
etc. up to key $2^k, k \in \mathbb{N}$.
This tagging makes it possible to select different parts of the entire
dataset in a uniform way, and perform queries of various thematic
selectivities. For example, if we selected all points of interest tagged with
key $1$, we would select all available points of interest, if we
selected all points of interest tagged with key  $2$, we would select half of
them, etc.

Every feature has a spatial extent as well that is modelled using the \linebreak GeoSPARQL
vocabulary. The spatial extent of the land ownership dataset constitutes a
uniform grid of $n \times n$ hexagons. The land ownership dataset forms the
basis for the spatial extent of all generated datasets since the size of each
dataset is given relatively to the number $n$. By modifying the number of
hexagons along an axis, we produce datasets of arbitrary size. As we will see
in the following section, this enabled us to adjust the selectivity of the
spatial predicates appearing in queries in a uniform way too.

As in \cite{vespa}, the generated land ownership dataset consists of $n^2$
features with hexagonal spatial extent, where each hexagon is placed  uniformly on a $n
\times n$ grid. The cardinality of the land ownerships is $n^2$.
The generated state dataset consists of $(\frac{n}{3})^2$ features with
hexagonal spatial extent, where each hexagon is placed  uniformly on a
$\frac{n}{3} \times \frac{n}{3}$ grid. The cardinality of the state
geometries is $(\frac{n}{3})^2$.
The generated road dataset consists of $n$ features with sloping line gometries.
Half of the line geometries are roughly horizontal and the other half are
roughly vertical. Each line consists of $\frac{n}{2}+1$ line segments.
The cardinality of the road geometries is $n$.
The generated point of interest dataset consists of $n^2$ features with point
geometries which are uniformly placed on $n$ sloping, evenly spaced, parallel
lines. The cardinality of the point of interest geometries is $n^2$. In
Figure~\ref{fig:sytheticDatasetVisualization} we present a sample of the
generated geometries.

		\subsubsection{Queries.}
		\label{sec:micro-def}
		\begin{table*}[t]
  \begin{center}
  \vspace{-6em}
  \caption{Query templates for generating SPARQL queries corresponding to (a)
	  spatial selections, and (b) spatial joins.}
	  \label{tbl:syntheticQueryTemplates}	
	  \hspace{-1em}	  
	\subtable[]
	  {
	  		\begin{tabular}{lcr}
			\texttt{}\\
			\texttt{}\\
			\texttt{SELECT ?s}\\ 
			\texttt{WHERE \{}\\
			\texttt{?s ns:hasGeometry/ns:asWKT ?g.}\\ 
			\texttt{?s c:hasTag/ns:hasKey "THEMA".}\\ 
			\texttt{FILTER(FUNCTION(?g, "GEOM"))\}}\\ 
			\end{tabular}			
			\label{tbl:syntheticQueryTemplatesSelect}
		}
	  \subtable[]
	  {			
			\begin{tabular}{lcr}  			
			\texttt{SELECT ?s1 ?s2}\\ 
			\texttt{WHERE \{}\\
			\texttt{?s1 ns1:hasGeometry/ns1:asWKT ?g1.}\\
			\texttt{?s1 ns1:hasTag/ns1:hasKey "THEMA".}\\
			\texttt{?s2 ns2:hasGeometry/ns2:asWKT ?g2.}\\
			\texttt{?s2 ns2:hasTag/ns2:hasKey "THEMA'".}\\  
			\texttt{FILTER(FUNCTION(?g1, ?g2))\}}\\
			\end{tabular}
		\label{tbl:syntheticQueryTemplatesJoin}	
		}
  \end{center}
\end{table*}

The synthetic workload generator produces SPARQL queries corresponding
to spatial selection and spatial joins by instantiating the two query templates
presented in Table~\ref{tbl:syntheticQueryTemplates}.

The query template used for producing SPARQL queries corresponding to spatial
selections is identical to the query template used in
\cite{brodtRDF3Xgermanoi,iswc2012-strabon}. In this query template, parameter
\texttt{THEMA} is one of the values used when assigning tags to a feature and
parameter \texttt{GEOM} is the WKT serialization of a rectangle.
As in \cite{iswc2012-strabon}, we define the \textit{thematic selectivity} of
an instantiation of the query template as the fraction of the total features of a
dataset that are tagged with a key equal to \texttt{THEMA}. For example, by
altering the value of \texttt{THEMA} from 1 to 2, we reduce the thematic
selectivity of the query by selecting half the nodes we previously did.  
We define the \textit{spatial selectivity} of an instantiation of the
query template as the fraction of the total features for which the topological
relations defined by parameter \texttt{FUNCTION} holds between each of them and the rectangle
defined by parameter \texttt{GEOM}. By modifying the value of the parameter namespace \texttt{ns} we
specify the dataset and the corresponding type of geometric information that
is examined by an instance of the query template.

The query template used for producing SPARQL queries corresponding to spatial
joins involves two datasets identified by the values of the parameter namespaces
\texttt{ns1} and \texttt{ns2}. In this query template as well, parameters \texttt{THEMA}
and \texttt{THEMA'} control the thematic selectivity of the query. The value of parameter
\texttt{FUNCTION} defines the topological relation that must hold between
instances of the two datasets that are involved in an instance of the query
template. For example, by instantiating the query template (b) with the values
\texttt{poi} for \texttt{ns1}, \texttt{state} for \texttt{ns2}, \texttt{1} for
\texttt{THEMA}, \texttt{2} for \texttt{THEMA'} and \texttt{geof:sfWithin} for
\texttt{FUNCTION}, we get a SPARQL query that asks for all generated points of
interest that are inside half of the generated states.

These query templates allow us to generate SPARQL queries with great diversity
regarding their spatial and thematic selectivity, thus stressing the optimizers
of the geospatial RDF stores that we test and evaluating their effectiveness in
identifying efficient query plans.

\section{Benchmark Results}
\label{sec:benchmarkresults}
In this section we present the results of running Geographica against three open source RDF stores.
As we mentioned earlier, we chose to test the systems Strabon, Parliament and uSeekM that currently
provide support for a rich subset of GeoSPARQL and stSPARQL.

	\subsection{Experimental Setup}
	\label{sec:experimentalsetup}
	In this section we describe the setup of the experiments used to evaluate the
selected triple stores. The machine that was used to run the benchmark is
equipped with two Intel Xeon E5620 processors with 12MB L3 cache running at 2.4 GHz, 24 GB
of RAM and a RAID-5 disk array that consists of four disks. Each disk has 32 MB
of cache and its rotational speed is 7200 rpm.

Each query in the micro and the synthetic benchmark was run three
times on cold and warm caches. For warm caches, we ran each query once before measuring
the response time, in order to warm up the caches.
We measured the response time for each query posed by measuring the elapsed
time from submitting the query until a complete iteration over the
results had been completed. The response time of each query was measured and the
median of all measurements is reported. A timeout of one hour is set as a time
limit for all queries.
For the macro benchmark, we run each scenario many times (with different initialization
each time) for one hour without
cleaning the caches and we report the average time for a complete execution
of all queries defined in each scenario.
Strabon and uSeekM utilize Postgres enhanced with PostGIS as a spatially-enabled
relational backend. For these systems, we set up an instance of Postgres 9.2 with
PostGIS 2.0 and we tuned it to make better use of the system resources.

For every dataset of Geographica, a unique property is used to connect geometries with their serialization (e.g. the Corine Land Use/Land cover ontology defines the property \texttt{clc:asWKT}), and this property is defined as a subproperty of the property \texttt{geo:asWKT} that is defined by GeoSPARQL.
Parliament is able to identify and index a triple that represents the serialization of a geometric object only when the property \texttt{geo:asWKT} is used. 
As a result, the RDFS reasoning capabilities of Parliament have to be enabled so that it performs forward chaining during data loading and indexes the geometry using the spatial index as well.
Strabon and uSeekM do not perform any reasoning on the input data.

	\subsection{Real-World Workload}
		\subsubsection{Dataset Storage.}
		\label{sec:resultsRealStore}
		In this section we present the time required by each system to store
and index the datasets of the real-world workload. As shown in
Table~\ref{table:StoringTime}, Strabon requires a lot of time to store and index the real-world dataset.
Stabon heavily indexes the produced DBMS tables since the existence of named graphs
require the creation of additional multi-column indices, thus leading to increased indexing time.
uSeekM needs less time since it is based on the native repository of Sesame which is known to be the most efficient implementation of Sesame for average sized datasets.
Since Parliament provides RDFS inference capabilities, it is reasonably slower that uSeekM as it requires more time to perform forward chaining on the input dataset.

\begin{table}[t]
\begin{center}
\begin{tabular}{|c|c|c|c|c|c|c|}
\hline
\textbf{Workload} & \textbf{Strabon} & \textbf{uSeekM} & \textbf{Parliament} \\
\hline
 \textbf{Real-world} & 550 sec.  & 214 sec. & 250 sec.  \\
 \textbf{Synthetic} & 221 sec.  & 406 sec. & 462 sec.  \\
\hline
\end{tabular}
\end{center}
\caption{Storing times}
\label{table:StoringTime}
\end{table}

		\subsubsection{Micro Benchmark.}
		\label{sec:resultsRealMicro}
\begin{figure*}[t]
  \begin{center}
	\vspace{-6em}	
	  \subfigure[][Cold caches]
	  {\includegraphics{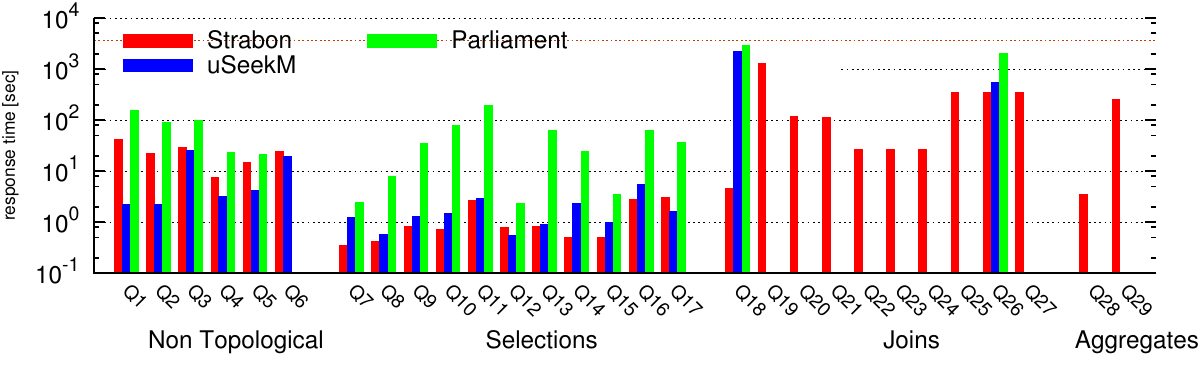}
	  \label{fig:RealCold}}
	  \hspace{-1em}
	  \\
	  \subfigure[][Warm caches]
	  {\includegraphics{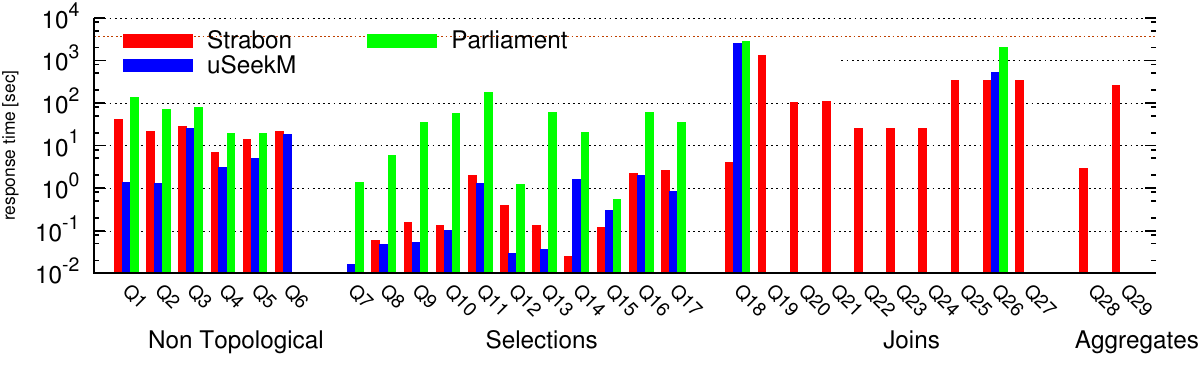}
	  \label{fig:RealWarm}}
	  \hspace{-1em}
	  \\
	\caption{Response times, real-world workload}
	\label{fig:RealResponseTimes}
  \end{center}
\end{figure*}

		The results of the micro benchmark are shown in Figure~\ref{fig:RealResponseTimes}
where the response time of each query is reported for both cold and warm caches.

First, results about computing non-topological functions are reported. For this
class of queries uSeekM performs the best.
Both uSeekM and Strabon are extensions of Sesame, but uSeekM extends the
native store of Sesame which is known to be more efficient, for small datasets, than Sesame
implementations on top of a DBMS, like Strabon,
thus the performance gain.
Computing the area of polygons (Query 6) was tested only in
uSeekM and Strabon since Parliament does not offer such functionality.
Finally, comparing the performance of all systems in cold and warm caches, we observe that
none of the RDF stores exploits the warm caches when evaluating topological functions.
This is because the non-topological functions used in this set of queries are
computationally intensive (especially when complex geometries are used) and the
time spent in CPU usage dominates I/O time.

Both uSeekM and Strabon utilize PostGIS for evaluating the spatial part of a
query. Thus the performance of uSeekM and Stabon is comparable regarding selections.
Strabon performs slightly better
than uSeekM in cold caches, but uSeekM improves its performance on warm caches
and outperforms Strabon.
Parliament, in general, needs an order of magnitude more time than Strabon and uSeekM.
It is interesting to have a closer look at Queries 14 and 15 which are semantically
equivalent. Both ask for points that
have a given distance from a given point. However, Query 14 creates the buffer
of a given point with radius $r$ and asks for points which are within this
buffer, while Query 15 asks for points that have distance less than $r$ from the
given point.
Parliament and uSeekM perform better in Query 15 which does not require buffer
computation than in Query 14. However, Strabon performs the same in both queries in cold caches while
in warm caches it answers Query 14 faster than the other two systems.

In the case of spatial joins, uSeekM and Parliament are able to evaluate only queries
18 and 27 given the time limit of one hour. 
Parliament evaluates separately graph patterns corresponding to different
graphs, produces the Cartesian product between them, and then applies the
spatial predicate to the result of the Cartesian product. This strategy is very
costly, thus Parliament is not able to answer any spatial join given the time limit. 
uSeekM does not utilize PostGIS in cases of  
joins, but only for spatial selections, so Strabon, which fully exploits
the query engine of PostGIS outperforms uSeekM.
In all cases, warm caches do not affect the response time of
the queries since a large number of intermediate results is produced.
Finally, spatial aggregations are tested. Strabon is the only system that
supports spatial aggregates so we do not have any comparison here. The only thing to notice then is that Query 28 which simply computes the
minimum bounding box that contains all geometries of the GAG dataset, is much faster than
Query 29 which computes the union of the same geometries.

		\subsubsection{Macro Benchmark.}
		\label{sec:resultsRealMacro}
		The results of the macro benchmark are shown in Table~\ref{table:MacroResults}.
In this table we report the average time needed for a complete iteration of all the
queries of each scenario. The Reverse Geocoding scenario has two queries which
use the function distance with a fixed limit. uSeekM performs the best in this scenario
while Strabon needs an order of magnitude more time. The Map Search and
Browsing scenario has one thematic query and two queries which select points
and lines in a given rectangle. Strabon and uSeekM have similar performance in
this scenario while Parliament needs an order of magnitude more time.
Finally, the Rapid Mapping for Fire Monitoring scenario is the most demanding
scenario. It comprises three queries that return lines and polygons that are
located inside a given rectangle, and two complex queries which include
spatial joins and construct new geometries (boundary and intersection) on the
fly. Only Strabon can serve this scenario since uSeekM and Parliament needed
more than an hour to evaluate the query RM6.

\begin{table}
\begin{center}
\begin{tabular}{|c|c|c|c|c|c|c|}
\hline
\textbf{Scenario} & \textbf{Strabon} & \textbf{uSeekM} & \textbf{Parliament} \\
\hline
 Reverse Geocoding & 65s & 0.77s & 2.6s  \\
\hline
 Map Search and Browsing & 0.9s & 0.6s & 22.2  \\
\hline
 Rapid Mapping for Fire Monitoring & 207.4s & - & - \\
\hline
\end{tabular}
\end{center}
\caption{Average Iteration times - Macro Scenarios}
\label{table:MacroResults}
\end{table}
	
	\subsection{Synthetic Workload}
	Let us now discuss representative experiments that we run using a synthetic
	workload that was produced using the generator presented in
	Section~\ref{sec:benchmark}.
	We generated a dataset by setting $n=512$ and $k=9$, where $n$ is the number
	used for defining the cardinalities of the generated geometries, and $k$ is
	the number used for defining the cardinalities of the generated tag values.
	This instantiation of the synthetic generator produces $262,144$ land
	ownership instances, $28,900$ states, $512$ roads and $262,144$ points of
	interest. Each feature is tagged with key $1$, every other feature with key
	$2$, etc. up to key $512$. The resulting dataset consists of 3,880,224 triples
	and its size is 745 MB.

		\subsubsection{Dataset Storage.}
		\label{sec:resultsSyntheticStore}
		Table~\ref{table:StoringTime} presents the time required by each system to
store and index the synthetic dataset. uSeekM requires more time than Strabon
for storing the dataset, since it stores it in a Sesame native store and then it
stores triples with geometric information at PostGIS as well. This overhead is
significant compared to the total time required for storing the dataset, but
leads to better response times in some cases. As we have already explained in Section~\ref{sec:resultsRealStore}, 
Parliament needs more time to store the synthetic dataset as well for the real-world workload
because it performs forward chaining on the input dataset.

		\subsubsection{Queries.}
		\label{sec:resultsSyntheticQueries}
\begin{landscape}
\begin{figure*}[f]
  \begin{center}
	\vspace{-6em}
	  \subfigure[][\shortstack{Intersects\\ \small{ tag 1, cold caches}}]
	  {\includegraphics{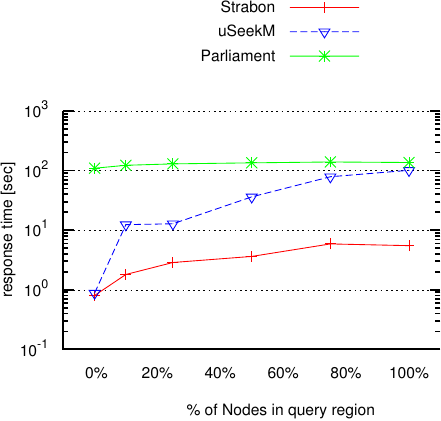}
	  \label{fig:SyntheticIntersects1Cold}}
	  \hspace{-1em}
	  \subfigure[][\shortstack{Intersects\\ \small{ tag 512, cold caches}}]
	  {\includegraphics{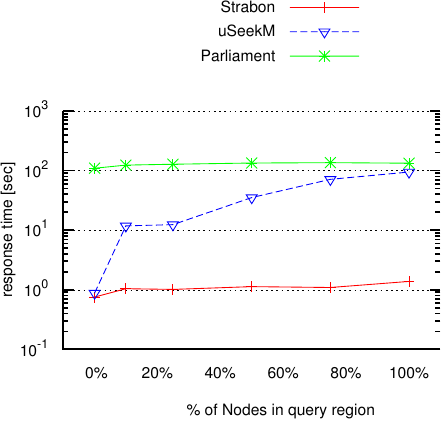}
	  \label{fig:SyntheticIntersects512Cold}}
	  \hspace{-1em}
	  \subfigure[][\shortstack{Within\\ \small{ tag 1, cold caches}}]
	  {\includegraphics{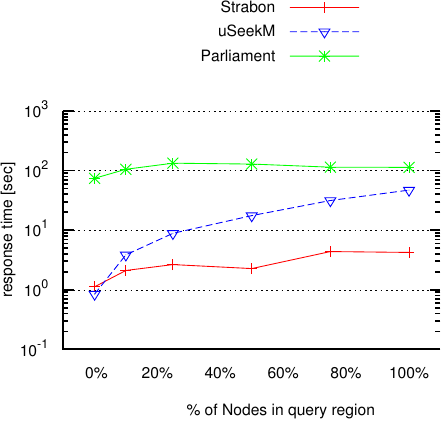}
	  \label{fig:SyntheticWithin1Cold}}
	  \hspace{-1em}
	  \subfigure[][\shortstack{Within\\ \small{tag 512, cold caches}}]
	  {\includegraphics{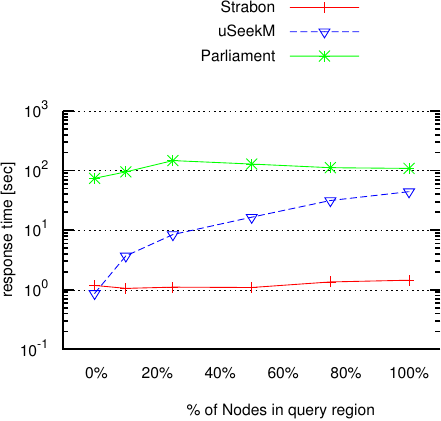}
	  \label{fig:SyntheticWithin512Cold}}
	  \hspace{-1em}
	  \\
	  \subfigure[][\shortstack{Intersects\\ \small{ tag 1, warm caches}}]
	  {\includegraphics{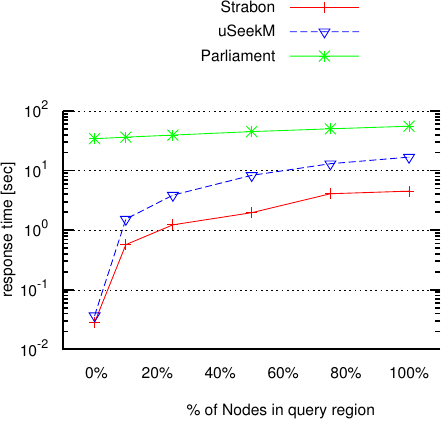}
	  \label{fig:SyntheticIntersects1Warm}}
	  \hspace{-1em}
	  \subfigure[][\shortstack{Intersects\\ \small{ tag 512, warm caches}}]
	  {\includegraphics{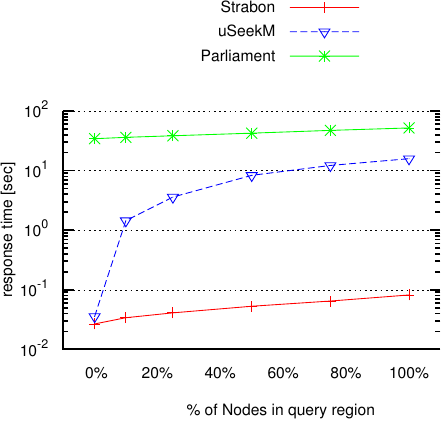}
	  \label{fig:SyntheticIntersects512Warm}}
	  \hspace{-1em}
	  \subfigure[][\shortstack{Within\\ \small{ tag 1, warm caches}}]
	  {\includegraphics{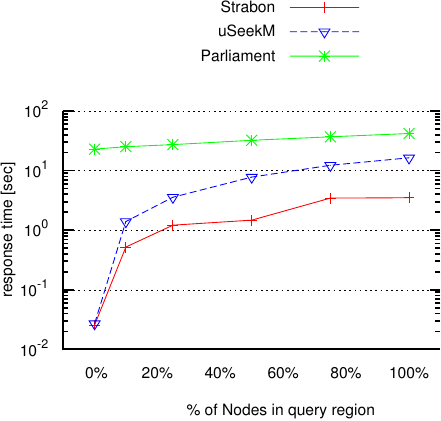}
	  \label{fig:SyntheticWithin1Warm}}
	  \hspace{-1em}
	  \subfigure[][\shortstack{Within\\ \small{ tag 512, warm caches}}]
	  {\includegraphics{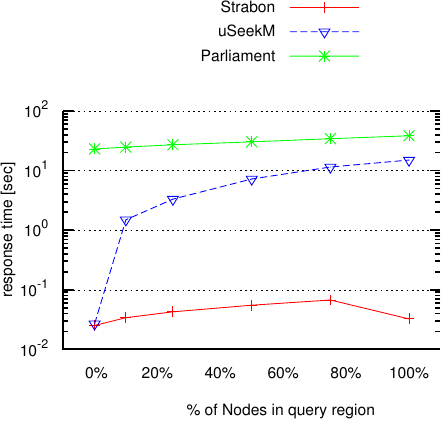}
	  \label{fig:SyntheticWithin512Warm}}
	  \hspace{-1em}
	  \\
	  \subfigure[][Intersects]
	  {\includegraphics[scale=0.62]{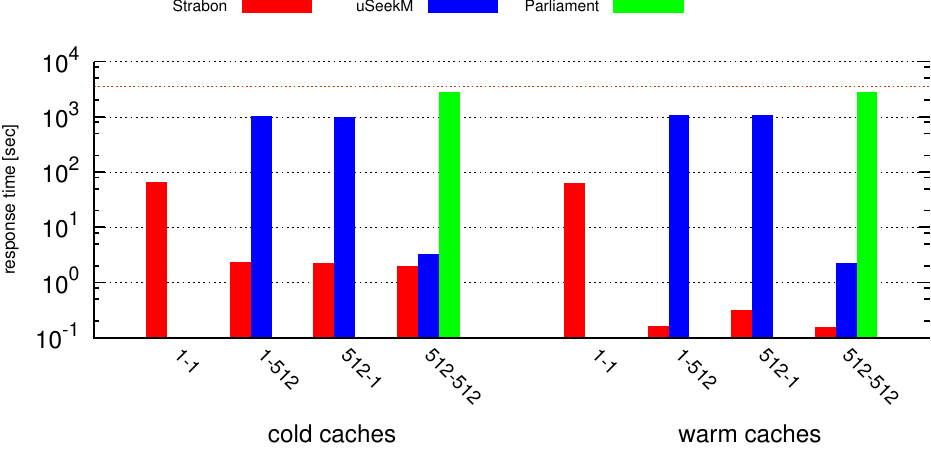}
	  \label{fig:SyntheticJoinIntersects}}	
	  \subfigure[][Touches]
	  {\includegraphics[scale=0.62]{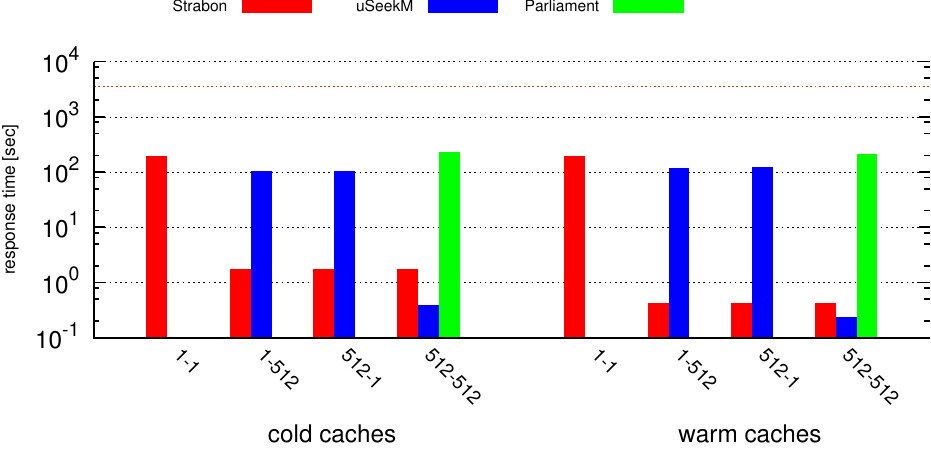}
	  \label{fig:SyntheticJoinTouches}}
	  \subfigure[][Within]
	  {\includegraphics[scale=0.62]{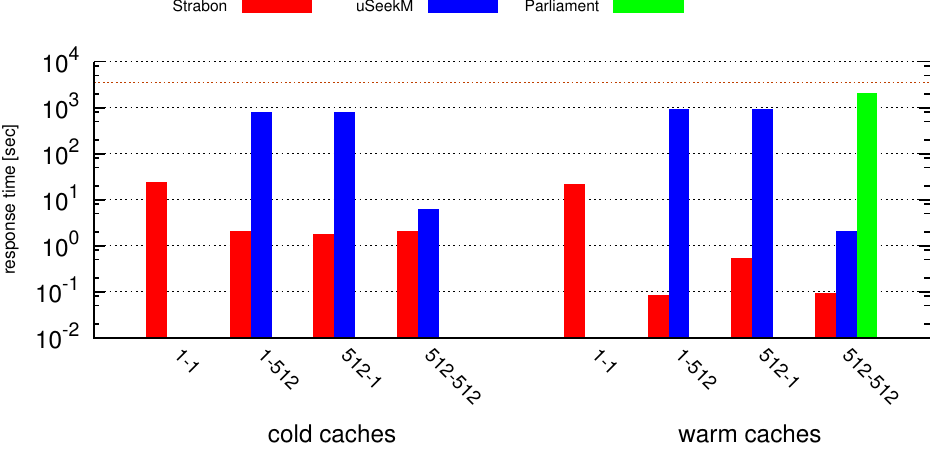}
	  \label{fig:SyntheticJoinWithin}}
	  \\
	\caption{Response times  - Synthetic Workload}
	\label{fig:SyntheticResponseTimes}
  \end{center}
\end{figure*}
\end{landscape}

		We instantiated the query template presented in
Table~\ref{tbl:syntheticQueryTemplatesSelect} in order to produce SPARQL
queries corresponding to spatial selections that ask for land ownerships that
intersect a given rectangle, and points of interest that are within a given
recangle. The given rectangle is generated in such as a way that the spatial
predicate of the query holds for 1\textpertenthousand, 10\%, 25\%, 50\%, 75\%
or all the features of the respective dataset. In addition, we instantiated
the query template using the extreme values \texttt{1} and \texttt{512} of the
parameter \texttt{THEMA} for selecting either all or approximatly
2\textpertenthousand  \mbox{ } of the total features  of a dataset.
The response time of each system for evaluating the instantiations of this query
template are presented in
Figures~\ref{fig:SyntheticIntersects1Cold}-\ref{fig:SyntheticWithin512Warm}.

We instantiated the query template presented in
Table~\ref{tbl:syntheticQueryTemplatesJoin} in order to produce SPARQL queries
corresponding to spatial joins that ask for land ownerships that intersect a
state, touching states and points of interest that are located inside a state.
We also instantiated this query template using all combinations of the extreme
values \texttt{1} and \texttt{512} for the parameters \texttt{THEMA} and
\texttt{THEMA'}.
The response time of each system for evaluating the instantiations of this query
template are presented in
Figures~\ref{fig:SyntheticJoinIntersects}-\ref{fig:SyntheticJoinWithin}.

By examining Figures~\ref{fig:SyntheticIntersects1Cold}-~\ref{fig:SyntheticWithin512Warm}, we observe that Strabon has very good performance overall. Strabon pushes the
evaluation of a SPARQL query to the underlying spatially-enabled DBMS, which in
this case is Postgres enhanced with PostGIS.
PostGIS has recently been enhanced with selectivity estimation capabilities. As
a result, when a query selects only a few geometries, query evaluation always
starts with the evaluation of the spatial predicate using the spatial index,
thus resulting in few intermmediate results and good response times. While the
spatial selectivity increases and more geometries satisfy the spatial predicate,
the optimizer of Postgres chooses different query plans. For example, when the
value of the parameter \texttt{THEMA} is $1$ (Figures
\ref{fig:SyntheticIntersects1Cold}, \ref{fig:SyntheticWithin1Cold},
\ref{fig:SyntheticIntersects1Warm}, \ref{fig:SyntheticWithin1Warm}) and the
value of the parameter \texttt{GEOM} is such that all geometries satisfy the spatial
predicate, Postgres ignores the spatial index and performs a sequential scan on
the table storing the geometries for evaluating the spatial predicate.
Similarly, when the value of the parameter \texttt{THEMA} is $512$ (Figures
\ref{fig:SyntheticIntersects512Cold}, \ref{fig:SyntheticWithin512Cold},
\ref{fig:SyntheticIntersects512Warm}, \ref{fig:SyntheticWithin512Warm})
and the value of the parameter \texttt{GEOM} is such that all
geometries satisfy the spatial predicate, Postgres starts with the evaluation of
the thematic selection that produces few intermediate results since only
2\textpertenthousand \mbox{ } of the features satisfy the thematic predicate,
resulting in good query response times.
In the case of spatial joins (Figures~\ref{fig:SyntheticJoinIntersects}-~\ref{fig:SyntheticJoinWithin}), Strabon is the fastest system in most cases.
The optimizer of Postgres takes into account the thematic selectivity of the
queries and selects good query plans, thus Strabon is the only system that is
able to answer the spatial joins given the one hour timeout when the parameters \texttt{THEMA} and \texttt{THEMA'} are equal to $1$.

Regarding uSeekM, we observe that its performance is not affected by the
thematic selectivity of the query. For spatial selections, uSeekM always start
by evaluating the spatial predicate in PostGIS and then continues the query
evaluation in the native Sesame store. As a result, regardless of the thematic
selectivity, the response time of uSeekM increases while increasing the number
of features with geometies that satisfy the given spatial predicate.

Regarding Parliament, we observe that its performance is not affected neither by
the thematic nor by the spatial selectivity of a query. Parliament always starts
by evaluating the non-spatial part of a query and then applies the thematic
filter and evaluates the spatial predicate exhaustively on the intermediate
results. Thus, the thematic and spatial selectivity of a query do not affect
the response time of Parliament.

In the case of spatial joins, uSeekM and Parliament produces the Cartesian
product between the graph patterns that are joined through the spatial
predicate, and evaluate the spatial predicate afterwards. This strategy is very
costly, thus Parliament is not able to answer most spatial joins given the
one hour timeout and uSeekM is more than two orders of magnitude slower than
Strabon.
However, in Figure~\ref{fig:SyntheticJoinTouches} we observe that uSeekM
outperforms Strabon. Strabon stores all geometries in a single table, so the
evaluation of the spatial predicate $Touches$ on this table returns not only the
geometries of states that touch each other, but the touching geometries of land
ownerships as well. The touching geometries of land ownerships are discarded
later on, but this overhead proves to be more costly than producing a
Cartesian product and evaluating the spatial predicate afterwards.

\section{Conclusions}
\label{sec:conclusions}
We presented a benchmark for evaluating the performance of geospatial RDF
stores that are beginning to emerge. 
We defined two workloads 
that test on the one hand the performance of the spatial component of such systems in isolation, 
and on the other hand test whether spatial query processing is deeply integrated in their
query engines.

\bibliographystyle{splncs03}
{\small
\bibliography{bibliography}
}

\end{document}